\documentclass[11pt,a4paper,english,nofootinbib]{revtex4}
\usepackage{lmodern}

\usepackage[T1]{fontenc}
\usepackage[latin9]{inputenc}
\usepackage{color}
\usepackage{babel}
\usepackage{amsmath}
\usepackage{amssymb}
\usepackage{esint}
\usepackage[unicode=true,pdfusetitle,
 bookmarks=true,bookmarksnumbered=false,bookmarksopen=false,
 breaklinks=false,pdfborder={0 0 1},backref=false,colorlinks=false]
 {hyperref}

\makeatletter

\pdfpageheight\paperheight
\pdfpagewidth\paperwidth

\@ifundefined{textcolor}{}
{%
 \definecolor{BLACK}{gray}{0}
 \definecolor{WHITE}{gray}{1}
 \definecolor{RED}{rgb}{1,0,0}
 \definecolor{GREEN}{rgb}{0,1,0}
 \definecolor{BLUE}{rgb}{0,0,1}
 \definecolor{CYAN}{cmyk}{1,0,0,0}
 \definecolor{MAGENTA}{cmyk}{0,1,0,0}
 \definecolor{YELLOW}{cmyk}{0,0,1,0}
}

\usepackage{latexsym}\usepackage{bm}

\makeatother

\definecolor{green}{RGB}{0, 180, 0}
\definecolor{cyan}{RGB}{0, 180, 180}
\definecolor{yellow}{RGB}{211,211,0}

\makeatother

\begin{document}

\title{Edge modes and Surface-Preserving Symmetries in Einstein-Maxwell Theory}

\author{Mohammad Reza Setare}
\email{rezakord@ipm.ir}

\affiliation{{Department of Science, Campus of Bijar, University of Kurdistan, Bijar, Iran }}

\author{Hamed Adami}
\email{hamed.adami@yahoo.com}

\affiliation{{Research Institute for Astronomy and Astrophysics of Maragha (RIAAM),
P.O. Box 55134-441, Maragha, Iran }}

\begin{abstract}
Einstein-Maxwell theory is not only covariant under diffeomorphisms but also is under $U(1)$ gauge transformations. We introduce a combined transformation constructed out of diffeomorphism and $U(1)$ gauge transformation. We show that symplectic potential, which is defined in covariant phase space method, is not invariant under combined transformations. In order to deal with that problem, following Donnelly and Freidel proposal \cite{1}, we introduce new fields. In this way, phase space and consequently symplectic potential will be extended. We show that new fields produce edge modes. We consider surface-preserving symmetries and we show that the group of surface-preserving symmetries is semi-direct sum of 2-dimensional diffeomorphism group on a spacelike codimension two surface with $SL(2,\mathbb{R})$ and $U(1)$. Eventually, we deduce that the Casimir of $SL(2,\mathbb{R})$ is the area element, similar to the pure gravity case \cite{1}.
\end{abstract}

\maketitle

\section{Introduction}\label{S.I}
Boundary conditions play crucial role in modern theoretical physics, for example in the holographic principle \cite{14,15}, the AdS/CFT correspondence \cite{16,17,18}, the bulk-boundary correspondence of condensed matter \cite{19,20,21}, or the study of entropy \cite{22,23,24,25,26,27}. We know that consideration of gauge field theories on manifolds with boundaries leads to the breaking the gauge invariance at the region's boundary. By the introduction edge modes which are compensating fields at the boundary, one can restore gauge invariance fully. Considering these boundary degrees of freedom in quantum theory need to extend Hilbert space \cite{25,28,30,31,32}. But in the classical theory we need to extend the phase space. By extension of the phase space, the symplectic structure takes a boundary part which contain physical edge mode fields. In the paper \cite{1} the authors have considered the problem of defining localized subsystems in gauge theory and gravity. By introducing new degrees of freedom on the boundary they have presented a general formalism to associate a gauge-invariant classical phase space to a spatial slice with boundary. Following \cite{1}, Geiller \cite{12} has shown that a systematic way of identifying degrees of freedom and possible associated boundary observables can be achieved by extending the covariant Hamiltonian formalism. In another work \cite{13} Giller has constructed the extended phase space for three-dimensional gravity in first order formalism. He has studied the boundary symmetries and the integrability of their generators and has found that the infinite-dimensional algebra of boundary symmetries with first order variables is the same as that with metric variables.

In this paper we study edge modes and surface-preserving symmetries in Einstein-Maxwell theory in 4-dimension. We study a combined transformation constructed out of diffeomorphism and $U(1)$ gauge transformation. We use the covariant phase space method to obtain conserved charges. Since the symplectic potential is not invariant under mentioned combined transformations, we follow the approach of paper \cite{1} and introduce new fields. So we extend the phase space and consequently symplectic potential and show that new fields produce edge modes. Then we investigate surface-preserving symmetries and show that the group of surface-preserving symmetries is semi-direct sum of 2-dimensional diffeomorphism group on a spacelike codimension two surface with $SL(2,\mathbb{R})$ and $U(1)$.
\section{Preliminary Considerations}\label{S.II}
In this paper, we will focus on the Einstein-Maxwell theory in which dynamical fields are spacetime metric $g_{\mu \nu}$ and the $U(1)$ gauge field $A_{\mu}$.  Suppose spacetime $(\mathcal{M},g)$ is globally hyperbolic and orientable. Suppose $\Phi$ is collection of dynamical fields, i.e. $\Phi = \{ g_{\mu \nu} , A_{\mu} \}$. The action describing Einstein-Maxwell theory is
\begin{equation}\label{1}
  S= \int_{\mathcal{V}} L[\Phi] d^{4}x,
\end{equation}
where
\begin{equation}\label{2}
  L[\Phi]=\sqrt{-g} ( R-2 \Lambda - 4 \pi F_{\mu \nu} F^{\mu \nu}),
\end{equation}
and $\mathcal{V}$ is a submanifold of spacetime manifold $\mathcal{M}$, i.e. $\mathcal{V} \subseteq \mathcal{M} $. Here $R$, $F_{\mu \nu} = \partial _{\mu} A_{\nu} - \partial _{\nu} A_{\mu} $ and $\Lambda$ are respectively the Ricci scalar, electromagnetic field strength and the cosmological constant. The Lagrangian \eqref{2}, and consequently the action \eqref{1}, are invariant under both diffeomorphisms and $U(1)$ gauge transformations. Let $\text{Y} : \mathcal{M} \rightarrow \mathcal{M}$ be a diffeomorphism of spacetime and $\text{Y}^{*} \mathcal{T}$ denote the pullback under diffeomorphism, where $\mathcal{T}$ is a tensor density (see Appendix \ref{APP.A}). For instance, pullback of metric and $U(1)$ gauge field are
\begin{equation}\label{3}
  \text{Y}^{*} g_{\alpha \beta}(x) = \frac{ \partial \text{Y}^{\alpha^{\prime}}}{\partial x^{\alpha}} \frac{ \partial \text{Y}^{\beta^{\prime}}}{\partial x^{\beta}} g_{\alpha^{\prime} \beta^{\prime}}(\text{Y}),
\end{equation}
\begin{equation}\label{4}
  \text{Y}^{*} A_{\alpha}(x) = \frac{ \partial \text{Y}^{\alpha^{\prime}}}{\partial x^{\alpha}} A_{\alpha^{\prime}}(\text{Y}),
\end{equation}
respectively and diffeomorphism invariance of the Lagrangian means that
\begin{equation}\label{5}
  \text{Y}^{*} L[\Phi] = L[\text{Y}^{*} \Phi].
\end{equation}
Let $\mathfrak{g}^{*}$ denote the $U(1)$ gauge transformation generated by $\lambda$, which acts on metric and $U(1)$ gauge field as
\begin{equation}\label{6}
  \mathfrak{g}_{\lambda}^{*} g_{\alpha \beta}(x) =g_{\alpha \beta}(x),
\end{equation}
\begin{equation}\label{7}
  \mathfrak{g}_{\lambda}^{*} A_{\alpha}(x) =A_{\alpha}(x)+ \partial _{\alpha} \lambda (x) ,
\end{equation}
respectively and $U(1)$ gauge invariance of the Lagrangian means that
\begin{equation}\label{8}
  \mathfrak{g}_{\lambda}^{*} L[\Phi] = L[\mathfrak{g}_{\lambda}^{*} \Phi]= L[\Phi].
\end{equation}
First order variation of the action \eqref{1} is
\begin{equation}\label{9}
\begin{split}
   \delta S & = \int_{\mathcal{V}} \delta L [\Phi]d^{4}x \\
     &= \int_{\mathcal{V}}  ( E_{\Phi}[\Phi] \delta \Phi + \partial_{\mu} \Theta ^{\mu}[\Phi ; \delta \Phi])d^{4}x\\
     &= \int_{\mathcal{V}} E_{\Phi}[\Phi] \delta \Phi d^{4}x + \int_{\partial\mathcal{V}} \Theta ^{\mu}[\Phi ; \delta \Phi]  d^{3}x_{\mu},
\end{split}
\end{equation}
in which $E_{\Phi}$ have dual indices with $\Phi$ and sum on $\Phi$ is explicitly assumed. In the equation \eqref{9}, $\Theta ^{\mu}[\Phi , \delta \Phi]$ is the surface term. Also, $E_{\Phi}=0$ give us the field equations. In Einstein-Maxwell theory, they are given as
\begin{equation}\label{10}
  E_{(g)}^{\mu \nu}= - \sqrt{-g} \left( G^{\mu \nu} + \Lambda g^{\mu \nu} - 8 \pi T^{\mu \nu} \right)=0,
\end{equation}
\begin{equation}\label{11}
  E_{(A)}^{\mu}=16 \pi \sqrt{-g} \nabla _{\nu} F^{\nu \mu}=0,
\end{equation}
\begin{equation}\label{12}
  \Theta^{\mu}[\Phi; \delta \Phi]= 2 \sqrt{-g} \left\{ \nabla ^{[\alpha} \left( g^{\mu] \beta} \delta g_{\alpha \beta} \right) -8 \pi F^{\mu \nu} \delta A_{\nu} \right\}.
\end{equation}
Equations \eqref{10} are known as Einstein's field equations, where $G^{\mu \nu}$ is the Einstein tensor and $T^{\mu \nu }$ is the electromagnetic energy-momentum tensor
\begin{equation}\label{13}
  T^{\mu \nu} = F^{\mu \alpha} F^{\nu}_{\hspace{1.7 mm} \alpha} - \frac{1}{4} g^{\mu \nu} F^{\alpha \beta} F_{\alpha \beta}.
\end{equation}
Also, equations \eqref{11} together with $ \nabla_{[\lambda}F_{\mu \nu]}=0 $ are Maxwell field equations in the curved spacetime.\\
The covariant phase space method \cite{2,3,4,5,6} provides a Hamiltonian description so that we can study the generators of infinitesimal
gauge transformations without the need to resort to a non-manifestly-covariant decomposition between space and time. In this method, one works with the space $\mathcal{S}$ of solutions to the field equations and $\mathcal{S}$ can be used to construct a phase space. The treatment of the exterior calculus given in \cite{1} could be useful for describing the phase space symplectic geometry on the $\mathcal{S}$ (such an approach is discussed more precisely in Ref.\cite{7}). The variation $\delta$ is treated as exterior derivative on phase space and $\delta \Phi$ defines a 1-form on the $\mathcal{S}$. It is worth mentioning that $\delta \Phi$ is a solution to linearized field equations. Let $\hat{V}$ be a vector on $\mathcal{S}$ and $I_{\hat{V}}$ denote interior product in $\hat{V}$ hence the interior product of $\delta \Phi$ in $\hat{V}$, $I_{\hat{V}} \delta \Phi$, becomes a scalar on $\mathcal{S}$. The action of the $\mathcal{S}$ Lie derivative $\mathcal{L}_{\hat{V}}$ is related to $\delta $ and $I_{\hat{V}}$ via Cartan's magic formula
\begin{equation}\label{14}
  \mathcal{L}_{\hat{V}} = I_{\hat{V}}\delta +\delta I_{\hat{V}}.
\end{equation}
Taking exterior derivative from Eq.\eqref{9} on $\mathcal{S}$ and using nilpotency $\delta ^{2}=0$, one finds that
\begin{equation}\label{15}
  \delta \int_{\partial\mathcal{V}} \Theta ^{\mu}[\Phi ; \delta \Phi]  d^{3}x_{\mu} = - \int_{\mathcal{V}} \delta E_{\Phi}[\Phi] \delta \Phi d^{4}x.
\end{equation}
Here the wedge product on phase space is implicit. Let spacetime region $\mathcal{V}$ be bounded by two Cauchy surfaces, say $\Sigma_{1}$ and $\Sigma_{2}$, and a time-like hypersurface $\mathcal{B}$, i.e. $\partial \mathcal{V} = (-\Sigma_{1}) \cup \Sigma_{2} \cup \mathcal{B}$, where the minus sign in front of $\Sigma_{1}$ serves to remind us that while the normal to $\partial \mathcal{V}$ must be directed outward (the normal to $\Sigma_{1}$ is future-directed and therefore points inward). Now, Eq.\eqref{15} can be written as
\begin{equation}\label{16}
  \delta \int_{\Sigma_{2}} \Theta ^{\mu}[\Phi ; \delta \Phi]  d^{3}x_{\mu}-\delta \int_{\Sigma_{1}} \Theta ^{\mu}[\Phi ; \delta \Phi]  d^{3}x_{\mu} = - \int_{\mathcal{V}} \delta E_{\Phi}[\Phi] \delta \Phi d^{4}x-\delta \int_{\mathcal{B}} \Theta ^{\mu}[\Phi ; \delta \Phi]  d^{3}x_{\mu}.
\end{equation}
It is clear from Eq.\eqref{16} that the quantity $\delta \int_{\Sigma} \Theta ^{\mu}[\Phi , \delta \Phi]  d^{3}x_{\mu}$ is independent of choosing the Cauchy surface when the flow of symplectic current $\omega^{\mu}= \delta \Theta^{\mu}$ from the time-like hypersurface $\mathcal{B}$ vanishes as well as equations of motion and linearized equations of motion are satisfied by $\Phi$ and $\delta \Phi$, respectively. We can define symplectic potential as
\begin{equation}\label{17}
  \boldsymbol{\Theta}_{\Sigma} [\Phi ; \delta \Phi]= \int_{\Sigma} \Theta ^{\mu}[\Phi ; \delta \Phi]  d^{3}x_{\mu},
\end{equation}
and consequently pre-symplectic form as exterior derivative of symplectic potential
\begin{equation}\label{18}
  \Omega [\Phi ; \delta \Phi, \delta \Phi]= \delta \boldsymbol{\Theta}_{\Sigma} [\Phi ; \delta \Phi].
\end{equation}
Pre-symplectic form is a 2-form on $\mathcal{S}$. Solution phase space can be constructed by factoring out the degeneracy subspace of configuration space (see Ref.\cite{2} for detailed discussion). Hence $\Omega$ will be a symplectic form on solution phase space and it is closed, skew-symmetric and nondegenerate. $\boldsymbol{\Theta}_{\Sigma} [\Phi ; \delta \Phi]$ is not invariant under diffeomorphism and $U(1)$ gauge transformation. It appears as a problem because we would like to define $\boldsymbol{\Theta}_{\Sigma}$ as a one-form on the gauge orbits, and this is only possible if $\boldsymbol{\Theta}_{\Sigma}$ is gauge-invariant (here we consider diffeomorphism as a gauge transformation as well). It has been shown recently that restoration of gauge invariance requires new boundary degrees of freedom \cite{1,7,12,13}.
\section{Introducing combined transformation}\label{S.III}
Unlike pure gravity, in Einstein-Maxwell theory there exists $U(1)$ freedom in addition to diffeomorphism so that the describing action (equivalently, the Lagrangian) is invariant under the action of both of them. There are some evidence that these two transformations should be combined \cite{6,8,9,10,11}. For this purpose, we introduce a combined map as $\mathcal{E}(\text{Y}, \mathfrak{g}_{\lambda})$ so that it induces a transformation $\mathcal{E}(\text{Y}, \mathfrak{g}_{\lambda}) ^{*}= \text{Y}^{*} \mathfrak{g}_{\lambda}^{*}$ which acts on dynamical fields as follwos:
\begin{equation}\label{19}
\begin{split}
   \mathcal{E}(\text{Y}, \mathfrak{g}_{\lambda})^{*} g_{\alpha \beta}(x)  & = \text{Y}^{*} \mathfrak{g}_{\lambda}^{*} g_{\alpha \beta}(x)\\
     & =\text{Y}^{*} g_{\alpha \beta}(x),
\end{split}
\end{equation}
\begin{equation}\label{20}
\begin{split}
   \mathcal{E}(\text{Y}, \mathfrak{g}_{\lambda})^{*} A_{\alpha}(x)  & = \text{Y}^{*} \mathfrak{g}_{\lambda}^{*} A_{\alpha}(x)\\
     & =\text{Y}^{*} [A_{\alpha}(x)+ \partial_{\alpha} \lambda (x)],
\end{split}
\end{equation}
where equations \eqref{6} and \eqref{7} were used. Variation of the pullback of a generic tensor density $\mathcal{T}$ is given by (see Appendix \ref{APP.A})
\begin{equation}\label{21}
  \delta \text{Y}^{*}\mathcal{T}= \text{Y}^{*} ( \delta \mathcal{T}+ \pounds _{\Delta_{\text{Y}}} \mathcal{T} ),
\end{equation}
where $ \Delta_{\text{Y}} = \delta \text{Y} \circ \text{Y}^{-1}$. Then variation and $\mathcal{E}(\text{Y}, \mathfrak{g}_{\lambda})^{*}$ do not commute and for dynamical fields we have
\begin{equation}\label{22}
\begin{split}
   \delta \mathcal{E}(\text{Y}, \mathfrak{g}_{\lambda})^{*} g_{\alpha \beta}(x) & = \delta \text{Y}^{*} g_{\alpha \beta}\\
     & =\text{Y}^{*} (\delta g_{\alpha \beta}+\pounds _{\Delta_{\text{Y}}} g_{\alpha \beta}),
\end{split}
\end{equation}
\begin{equation}\label{23}
\begin{split}
   \delta \mathcal{E}(\text{Y}, \mathfrak{g}_{\lambda})^{*} A_{\alpha}(x)  & = \delta \text{Y}^{*} (A_{\alpha}+ \partial_{\alpha} \lambda )\\
     & =\text{Y}^{*} (\delta A_{\alpha}+ \partial_{\alpha} \delta\lambda +\pounds _{\Delta_{\text{Y}}} A_{\alpha}+ \partial_{\alpha} \pounds _{\Delta_{\text{Y}}}\lambda),
\end{split}
\end{equation}
Now we want to see that how the symplectic potential behaves under the action of $\mathcal{E}(\text{Y}, \mathfrak{g}_{\lambda})^{*}$. To this end, let us consider
\begin{equation}\label{24}
  \begin{split}
     \Theta^{\mu}[\mathcal{E}(\text{Y}, \mathfrak{g}_{\lambda})^{*} \Phi ;  \delta \mathcal{E}(\text{Y}, \mathfrak{g}_{\lambda})^{*} \Phi] = \text{Y}^{*} \{ & \Theta^{\mu} [\Phi ; \delta \Phi]+\Theta^{\mu} [\Phi ; \pounds_{\Delta_{\text{Y}}} g_{\alpha \beta} ,\pounds_{\Delta_{\text{Y}}} A_{\alpha} + \partial_{\alpha} (\delta \lambda + \pounds_{\Delta_{\text{Y}}} \lambda)] \} \\
     = \text{Y}^{*} \{ & \Theta^{\mu} [\Phi ; \delta \Phi]+ L[\Phi] \Delta_{\text{Y}}^{\mu} +\partial_{\nu} \Pi^{\mu \nu}[\Phi ,\text{Y}, \lambda ; \delta \text{Y}, \delta \lambda ]\\
       & - 2 E_{(g) \nu}^{\mu} \Delta_{\text{Y}}^{\nu} - E_{(A)}^{\mu} (\delta \lambda + \pounds_{\Delta_{\text{Y}}} \lambda + A_{\alpha} \Delta_{\text{Y}}^{\alpha} )\}.
  \end{split}
\end{equation}
This is not simply the pullback of the surface term even for the on-shell case. In the equation \eqref{24},
\begin{equation}\label{25}
  \Pi^{\mu \nu}[\Phi ,\text{Y}, \lambda ; \delta \text{Y}, \delta \lambda ] = \Pi_{(g)}^{\mu \nu}(\Delta_{\text{Y}})+\Pi_{(A)}^{\mu \nu}(A_{\alpha} \Delta_{\text{Y}}^{\alpha}) +\Pi_{(A)}^{\mu \nu}(\delta \lambda)+\Pi_{(A)}^{\mu \nu}(\pounds_{\Delta_{\text{Y}}} \lambda)
\end{equation}
is an anti-symmetric tensor density of weight $+1$, where
\begin{equation}\label{26}
  \Pi_{(g)}^{\mu \nu}(\Delta_{\text{Y}})=-2 \sqrt{-g} \nabla^{[\mu} \Delta_{\text{Y}}^{\nu ]}, \hspace{1 cm} \Pi_{(A)}^{\mu \nu}( \lambda)=-16 \pi \sqrt{-g} F^{\mu \nu} \lambda .
\end{equation}
The symplectic potential $\boldsymbol{\Theta}_{\Sigma}$ is obtained by integrating the surface term $\Theta^{\mu}$ over the Cauchy surface $\Sigma$. $\boldsymbol{\Theta}_{\Sigma}$ is not invariant under both diffeomorphism and $U(1)$ gauge transformation, instead we have
\begin{equation}\label{27}
  \begin{split}
     \boldsymbol{\Theta}_{\Sigma} [\mathcal{E}(\text{Y}, \mathfrak{g}_{\lambda})^{*} \Phi ;  \delta \mathcal{E}(\text{Y}, \mathfrak{g}_{\lambda})^{*} \Phi]  &= \int_{\Sigma} \Theta^{\mu}[\mathcal{E}(\text{Y}, \mathfrak{g}_{\lambda})^{*} \Phi ;  \delta \mathcal{E}(\text{Y}, \mathfrak{g}_{\lambda})^{*} \Phi] d^{3}x_{\mu} \\
       & =\int_{\Sigma} \text{Y}^{*} \{ \Theta^{\mu} [\Phi ; \delta \Phi]+ L[\Phi] \Delta_{\text{Y}}^{\mu} +\partial_{\nu} \Pi^{\mu \nu}[\Phi ,\text{Y}, \lambda ; \delta \text{Y}, \delta \lambda ]\\
       &  \hspace{1.7 cm}- 2 E_{(g) \nu}^{\mu} \Delta_{\text{Y}}^{\nu} - E_{(A)}^{\mu} (\delta \lambda + \pounds_{\Delta_{\text{Y}}} \lambda + A_{\alpha} \Delta_{\text{Y}}^{\alpha} )\} d^{3}x_{\mu}\\
       & =\int_{\text{Y}(\Sigma)} \{ \Theta^{\mu} [\Phi ; \delta \Phi]+ L[\Phi] \Delta_{\text{Y}}^{\mu} +\partial_{\nu} \Pi^{\mu \nu}[\Phi ,\text{Y}, \lambda ; \delta \text{Y}, \delta \lambda ]\\
       &  \hspace{1.7 cm}- 2 E_{(g) \nu}^{\mu} \Delta_{\text{Y}}^{\nu} - E_{(A)}^{\mu} (\delta \lambda + \pounds_{\Delta_{\text{Y}}} \lambda + A_{\alpha} \Delta_{\text{Y}}^{\alpha} )\} d^{3}x_{\mu}
  \end{split}
\end{equation}
Now we need to introduce new variables into the phase space whose transformation laws will cancel the extra terms in the symplectic potential. Following \cite{1}, consider a coordinate system $\text{X}$ which we view as a mapping $\text{X}: U \rightarrow \mathcal{M}$ where $ U \subset \mathbb{R}^{4}$ is an open set and assume that $\text{X}$ is invertible on its image. We assume for simplicity that $\Sigma$ can be covered with one open set $\sigma \subset U$ so that under this map we have $\Sigma = \text{X}(\sigma)$. Under a diffeomorphism $\text{Y}: \mathcal{M} \rightarrow \mathcal{M}$, the coordinate system $\text{X}$ changes to
\begin{equation}\label{28}
  \bar{\text{X}} \equiv \text{Y}^{*} (\text{X})= \text{Y}^{-1} \circ \text{X}.
\end{equation}
 Therefore, we can define the symplectic potential $\boldsymbol{\Theta}_{\Sigma}[\Phi; \delta \Phi]$ as the integral over the slice $\Sigma = \text{X}(\sigma)$ so that under a diffeomorphism it transforms as
\begin{equation}\label{29}
\begin{split}
   \boldsymbol{\Theta}_{(\text{Y}^{-1} \circ \text{X})(\sigma)} [\mathcal{E}(\text{Y}, \mathfrak{g}_{\lambda})^{*} \Phi ;  \delta \mathcal{E}(\text{Y}, \mathfrak{g}_{\lambda})^{*} \Phi] &=\int_{\Sigma} \{ \Theta^{\mu} [\Phi ; \delta \Phi]+ L[\Phi] \Delta_{\text{Y}}^{\mu} +\partial_{\nu} \Pi^{\mu \nu}[\Phi ,\text{Y}, \lambda ; \delta \text{Y}, \delta \lambda ]\\
       &  \hspace{1.7 cm}- 2 E_{(g) \nu}^{\mu} \Delta_{\text{Y}}^{\nu} - E_{(A)}^{\mu} (\delta \lambda + \pounds_{\Delta_{\text{Y}}} \lambda + A_{\alpha} \Delta_{\text{Y}}^{\alpha} )\} d^{3}x_{\mu}.
\end{split}
\end{equation}
Consider the variation of dynamical fields
\begin{equation}\label{30}
  \begin{split}
     \delta \Phi & = \delta \text{Y}^{*}(\text{Y}^{-1})^{*} \Phi \\
       & = \text{Y}^{*} \{ \delta (\text{Y}^{-1})^{*} \Phi +\pounds_{\Delta_{\text{Y}}} (\text{Y}^{-1})^{*} \Phi \} \\
       & = \text{Y}^{*} (\text{Y}^{-1})^{*} \{ \delta \Phi +\pounds_{\Delta_{\text{Y}^{-1}}} \Phi  +\pounds_{\text{Y}^{*}\Delta_{\text{Y}}} \Phi \} \\
       & = \delta \Phi +\pounds_{\Delta_{\text{Y}^{-1}}} \Phi  +\pounds_{\text{Y}^{*}\Delta_{\text{Y}}} \Phi,
  \end{split}
\end{equation}
then we find that
\begin{equation}\label{31}
   \Delta_{\text{Y}^{-1}}^{\mu} = - \text{Y}^{*}\Delta_{\text{Y}}^{\mu}.
\end{equation}
Let $\bar{\text{X}}^{*} \Phi$ denote pullback of dynamical fields induced by $\bar{\text{X}}$ then
\begin{equation}\label{32}
  \begin{split}
    \delta  \bar{\text{X}}^{*} \Phi  &= \delta  (\text{Y}^{-1} \circ \text{X})^{*} \Phi \\
       & =\delta \text{X}^{*} (\text{Y}^{-1})^{*} \Phi  \\
       & = \text{X}^{*} \{ \delta (\text{Y}^{-1})^{*} \Phi + \pounds_{\Delta_{\text{X}}} (\text{Y}^{-1})^{*} \Phi \} \\
       & = \text{X}^{*} (\text{Y}^{-1})^{*} \{ \delta \Phi +\pounds_{\Delta_{\text{Y}^{-1}}} \Phi + \pounds_{\text{Y}^{*}\Delta_{\text{X}}} \Phi \}\\
       &=  \bar{\text{X}}^{*} \{ \delta \Phi -\pounds_{ \text{Y}^{*}\Delta_{\text{Y}}} \Phi + \pounds_{\text{Y}^{*}\Delta_{\text{X}}} \Phi \},
  \end{split}
\end{equation}
where in the last line Eq.\eqref{31} was used. We expect that the variation of $\bar{\text{X}}^{*} \Phi$ obey from Eq.\eqref{21} therefore we find that
\begin{equation}\label{33}
  \Delta_{\bar{\text{X}}}= \text{Y}^{*} (\Delta_{\text{X}}-\Delta_{\text{Y}}).
\end{equation}
This equation ensures that the presence of $\text{X}$ makes symplectic potential diffeomorphism invariant. Now just consider $U(1)$ gauge transformation. In order to make $\boldsymbol{\Theta}_{\Sigma}$ to be $U(1)$ gauge invariant we has to replace $\lambda$ in Eq.\eqref{29} by new scalar field $\Xi$ so that it transforms as
\begin{equation}\label{34}
  \Xi \rightarrow \bar{\Xi} =\Xi - \lambda
\end{equation}
under $U(1)$ gauge transformation \cite{12}. Now we are ready to extend phase space by introducing new fields $\text{X}$ and $\Xi$ which will be our task in the next section.
\section{Extension of phase space}\label{S.IV}
As mentioned earlier we can use Donnelly and Freidel proposal \cite{1} to extend phase space. Let $v$ be preimage of the subregion $\mathcal{V} \subset \mathcal{M}$ and $\text{X}^{*}$ denote the pullback induced by coordinates transformation $\text{X}$. Assume for simplicity that $\mathcal{V}$ can be covered with one open set $v \subset U$ so that under this map we have $\mathcal{V}= \text{X}(v)$. Therefore the action \eqref{1} can be written as
\begin{equation}\label{35}
  \begin{split}
     S & = \int_{\text{X}(v)} L[\Phi] d^{4}x\\
       & = \int_{v} \text{X}^{*} L[\Phi] d^{4}x \\
       & = \int_{v} \mathcal{E}(\text{X}, \mathfrak{g}_{\Xi})^{*} L[\Phi] d^{4}x \\
       & = \int_{v} L[ \mathcal{E}(\text{X}, \mathfrak{g}_{\Xi})^{*}\Phi] d^{4}x
  \end{split}
\end{equation}
where equations \eqref{5} and \eqref{8} were used. It is a minimal prescription for introducing new fields $\text{X}$ and $\Xi$ into the theory. Taking the variation of the action \eqref{35} leads to
\begin{equation}\label{36}
\begin{split}
   \delta S & = \int_{v} \delta L [\mathcal{E}(\text{X}, \mathfrak{g}_{\Xi})^{*} \Phi]d^{4}x \\
     &= \int_{v}  ( E_{\Phi}[\mathcal{E}(\text{X}, \mathfrak{g}_{\Xi})^{*} \Phi] \delta \mathcal{E}(\text{X}, \mathfrak{g}_{\Xi})^{*} \Phi + \partial_{\mu} \Theta ^{\mu}[\mathcal{E}(\text{X}, \mathfrak{g}_{\Xi})^{*} \Phi ; \delta \mathcal{E}(\text{X}, \mathfrak{g}_{\Xi})^{*} \Phi])d^{4}x .
\end{split}
\end{equation}
Because equations of motion are covariant under both diffeomorphism and $U(1)$ gauge transformations then using equations \eqref{22} and \eqref{23} we have
\begin{equation}\label{37}
  \begin{split}
     E_{\Phi}[\mathcal{E}(\text{X}, \mathfrak{g}_{\Xi})^{*} \Phi] \delta \mathcal{E}(\text{X}, \mathfrak{g}_{\Xi})^{*} \Phi & = \\
      \text{X}^{*}& \{ E_{(g)}^{\mu \nu} (\delta g_{\mu \nu}+\pounds _{\Delta_{\text{X}}} g_{\mu \nu})+ E_{(A)}^{\mu} (\delta A_{\mu}+ \partial_{\mu} \delta\Xi +\pounds _{\Delta_{\text{X}}} A_{\mu}+ \partial_{\mu} \pounds _{\Delta_{\text{X}}}\Xi) \} \\
       & =\text{X}^{*} \{ E_{\Phi}[\Phi] \delta \Phi + \partial _{\mu} [ 2 E_{(g) \nu}^{\mu} \Delta_{\text{X}}^{\nu} + E_{(A)}^{\mu} (\delta \Xi + \pounds_{\Delta_{\text{X}}} \Xi + A_{\alpha} \Delta_{\text{X}}^{\alpha} )] \}
  \end{split}
\end{equation}
where Bianchi identities $\nabla_{\nu} G^{\mu \nu}=0$ and $\nabla_{[\lambda} F_{\mu \nu ]}=0$ were used. By substituting Eq.\eqref{37} into Eq.\eqref{36} and using Eq.\eqref{24} we find that
\begin{equation}\label{38}
\begin{split}
   \delta S & = \int_{v}  \text{X}^{*}( E_{\Phi}[ \Phi] \delta \Phi + \partial_{\mu} \{ \Theta ^{\mu}[ \Phi ; \delta  \Phi] + L[\Phi] \Delta_{\text{X}}^{\mu} +\partial_{\nu} \Pi^{\mu \nu}[\Phi ,\text{X}, \Xi ; \delta \text{X}, \delta \Xi ] \})d^{4}x \\
   &= \int_{\mathcal{V}}  E_{\Phi}[ \Phi] \delta \Phi d^{4}x + \int_{\partial \mathcal{V}}  \{ \Theta ^{\mu}[ \Phi ; \delta  \Phi] + L[\Phi] \Delta_{\text{X}}^{\mu} +\partial_{\nu} \Pi^{\mu \nu}[\Phi ,\text{X}, \Xi ; \delta \text{X}, \delta \Xi ] \} d^{3}x_{\mu}.
\end{split}
\end{equation}
By introducing new fields the bulk term does not change and therefore the original equations of motion. In other words, new fields do not affect the dynamics of theory and they are not dynamical. Instead, the surface term has changed so that new expression for surface term is
\begin{equation}\label{39}
 \tilde{\Theta} ^{\mu}[ \Psi ; \delta  \Psi]= \Theta ^{\mu}[ \Phi ; \delta  \Phi] + L[\Phi] \Delta_{\text{X}}^{\mu} +\partial_{\nu} \Pi^{\mu \nu}[\Psi ; \delta \Psi ] ,
\end{equation}
where new fields are taken into account. Here $\Psi$ is a collection of fields so that, in addition to the dynamical fields $\Phi$, it contains new fields $\text{X}$ and $\Xi$. Extended symplectic potential $\tilde{\boldsymbol{\Theta}}[ \Psi ; \delta  \Psi]$ is obtained by integrating the new surface term $\tilde{\Theta} ^{\mu}[ \Psi ; \delta  \Psi]$ over the Cauchy surface $\Sigma$. Therefore, we define extended symplectic potential as
\begin{equation}\label{40}
  \begin{split}
     \tilde{\boldsymbol{\Theta}}[ \Psi ; \delta  \Psi] & = \int_{\sigma} \text{X}^{*} \tilde{\Theta} ^{\mu}[ \Psi ; \delta  \Psi] d^{3}x_{\mu} \\
     & = \int_{\Sigma} \tilde{\Theta} ^{\mu}[ \Psi ; \delta  \Psi] d^{3}x_{\mu} \\
     & = \int_{\Sigma} ( \Theta ^{\mu}[ \Phi ; \delta  \Phi] + L[\Phi] \Delta_{\text{X}}^{\mu} ) d^{3}x_{\mu}+ \int_{\partial \Sigma } \Pi^{\mu \nu}[\Psi ; \delta \Psi ] d^{2}x_{\mu \nu} .
  \end{split}
\end{equation}
Equation \eqref{40} differs from the symplectic potential for the non-extended phase space \eqref{17} by both a boundary term as well as a bulk term
coming from the on-shell value of the Lagrangian.\\
It is worth mentioning that new fields $\text{X}$ and $\Xi$ transform as
\begin{equation}\label{41}
  \bar{\text{X}}=\mathcal{E}(\text{Y},\mathfrak{g}_{\lambda})^{*} \text{X}= \text{Y}^{*}( \text{X}),
\end{equation}
\begin{equation}\label{42}
  \bar{\Xi}= \mathcal{E}(\text{Y},\mathfrak{g}_{\lambda})^{*} \Xi= \text{Y}^{*} (\Xi -\lambda),
\end{equation}
under combined transformation $\mathcal{E}(\text{Y},\mathfrak{g}_{\lambda})$, respectively. Now let us consider behaviour of the extended symplectic potential \eqref{40} under combined transformation
\begin{equation}\label{43}
  \begin{split}
       \tilde{\boldsymbol{\Theta}}[ \mathcal{E}(\text{Y},\mathfrak{g}_{\lambda})^{*}\Psi ; \delta  \mathcal{E}(\text{Y},\mathfrak{g}_{\lambda})^{*}\Psi] & = \int_{\mathcal{E}(\text{Y},\mathfrak{g}_{\lambda})^{*}\text{X}(\sigma)} \tilde{\Theta} ^{\mu}[ \mathcal{E}(\text{Y},\mathfrak{g}_{\lambda})^{*}\Psi ; \delta  \mathcal{E}(\text{Y},\mathfrak{g}_{\lambda})^{*}\Psi] d^{3}x_{\mu}  \\
       &  = \int_{\bar{\text{X}}(\sigma)} \bigl\{\Theta ^{\mu}[ \mathcal{E}(\text{Y},\mathfrak{g}_{\lambda})^{*}\Phi ; \delta  \mathcal{E}(\text{Y},\mathfrak{g}_{\lambda})^{*}\Phi] + L[\mathcal{E}(\text{Y},\mathfrak{g}_{\lambda})^{*}\Phi] \Delta_{\bar{\text{X}}}^{\mu}\\
       & \hspace{1.5 cm}+\partial_{\nu} \Pi^{\mu \nu}[\mathcal{E}(\text{Y},\mathfrak{g}_{\lambda})^{*} \Psi ; \delta \mathcal{E}(\text{Y},\mathfrak{g}_{\lambda})^{*}\Psi ] \bigr\} d^{3}x_{\mu} \\
       & = \tilde{\boldsymbol{\Theta}}[ \Psi ; \delta \Psi]- \int_{\Sigma} \{ 2 E_{(g) \nu}^{\mu} \Delta_{\text{Y}}^{\nu} + E_{(A)}^{\mu} (\delta \lambda + \pounds_{\Delta_{\text{Y}}} \lambda + A_{\alpha} \Delta_{\text{Y}}^{\alpha} ) \} d^{3}x_{\mu}.
  \end{split}
\end{equation}
It is clear from the above equation that the extended symplectic potential is invariant under combined transformations at least on-shell or for transformations which do not depend on the solution.
\section{Revisiting covariant phase space method}\label{S.V}
We have extended phase space so that $\Psi$ represent a point on the extended phase space. $\delta \Psi$ at a point of extended phase space  $\tilde{\mathcal{S}}$ describes an infinitesimal displacement away from a particular solution. Let $\xi^{\mu}(x)$ and $\lambda(x)$ be generators of diffeomorphism and $U(1)$ gauge transformation. We can introduce a combined transformation so that $\chi= (\xi , \lambda)$ is the generator of such transformations \cite{8}. $\chi= (\xi , \lambda)$ defines a vector field on $\mathcal{S}$, which we denote $\hat{\chi}$, whose action
on $\delta \Phi$ is
\begin{equation}\label{44}
  I_{\hat{\chi}} \delta \Phi= \pounds_{\xi} \Phi + \delta_{\Phi}^{A} \partial _{\mu} \lambda,
\end{equation}
where $\delta_{\Phi}^{\Phi^{\prime}}$ is Kronecker delta on collection of dynamical fields $\Phi= \{ g_{\mu \nu} , A_{\mu} \}$. Therefore, the action of the $\mathcal{S}$ Lie derivative along $\hat{\chi}$ can be defined via Cartan's magic formula $\mathcal{L}_{\hat{\chi}} = I_{\hat{\chi}}\delta +\delta I_{\hat{\chi}}$.\\
As we mentioned earlier, new fields are introduced via replacing dynamical fields by $\mathcal{E}(\text{X}, \mathfrak{g}_{\Xi})^{*} \Phi$ in the lagrangian. Because $\mathcal{E}(\text{X}, \mathfrak{g}_{\Xi})^{*} \Phi$ is invariant under combined transformation, i.e.
\begin{equation}\label{45}
\begin{split}
   \bar{\mathcal{E}}(\text{X}, \mathfrak{g}_{\Xi})^{*} \bar{\Phi} & =\bar{\mathcal{E}}(\text{X}, \mathfrak{g}_{\Xi})^{*} \mathcal{E}(\text{Y}, \mathfrak{g}_{\lambda})^{*} \Phi \\
   &= \bar{\text{X}}^{*} \mathfrak{g}_{\Xi-\lambda}^{*} \text{Y}^{*} \mathfrak{g}_{\lambda}^{*} \Phi  \\
     & = \bar{\text{X}}^{*} \text{Y}^{*} \mathfrak{g}_{\Xi} \Phi\\
     & = \text{X}^{*} (\text{Y}^{-1})^{*} \text{Y}^{*} \mathfrak{g}_{\Xi} \Phi\\
     & =  \mathcal{E}(\text{X}, \mathfrak{g}_{\Xi})^{*} \Phi,
\end{split}
\end{equation}
then its $\mathcal{S}$ Lie derivative along $\hat{\chi}$ vanishes\footnote{Where equations \eqref{28} and \eqref{34} were used}. Let consider the action of $\mathcal{S}$ Lie derivative on $\mathcal{E}(\text{X}, \mathfrak{g}_{\Xi})^{*} \Phi$ along $\hat{\chi}$
\begin{equation}\label{46}
  \begin{split}
     0 & = \mathcal{L}_{\hat{\chi}} \mathcal{E}(\text{X}, \mathfrak{g}_{\Xi})^{*} \Phi \\
       & = I_{\hat{\chi}} \delta \{ \text{X}^{*}  (\Phi+ \delta_{\Phi}^{A} \partial _{\mu} \Xi) \} \\
       & = \text{X}^{*} \{ I_{\hat{\chi}} \delta \Phi+ \delta_{\Phi}^{A} \partial _{\mu} I_{\hat{\chi}} \delta \Xi + \pounds _{ I_{\hat{\chi}} \Delta_{\text{X}}} \Phi+ \delta_{\Phi}^{A} \partial _{\mu} \pounds _{ I_{\hat{\chi}} \Delta_{\text{X}}} \Xi \}\\
       & = \text{X}^{*} \{ \pounds_{\xi} \Phi + \delta_{\Phi}^{A} \partial _{\mu} \lambda+ \delta_{\Phi}^{A} \partial _{\mu} I_{\hat{\chi}} \delta \Xi + \pounds _{ I_{\hat{\chi}} \Delta_{\text{X}}} \Phi+ \delta_{\Phi}^{A} \partial _{\mu} \pounds _{ I_{\hat{\chi}} \Delta_{\text{X}}} \Xi \}
  \end{split}
\end{equation}
This equation will be held when we have
\begin{equation}\label{47}
  I_{\hat{\chi}} \Delta_{\text{X}}^{\mu} = -\xi^{\mu}
\end{equation}
\begin{equation}\label{48}
  I_{\hat{\chi}} \delta \Xi = \pounds_{\xi} \Xi - \lambda .
\end{equation}
These two equations are infinitesimal versions of equations \eqref{41} and \eqref{42}. We can easily confirm from Eqs.\eqref{46}-\eqref{48} that $\delta \mathcal{E}(\text{X}, \mathfrak{g}_{\Xi})^{*} \Phi$ are annihilated by infinitesimal combined transformation generated by $\hat{\chi}$. Also, one can show that the extended simplectic form is annihilated on-shell by infinitesimal combined transformation generated by $\hat{\chi}$
\begin{equation}\label{49}
  \begin{split}
     I_{\hat{\chi}} \tilde{\boldsymbol{\Theta}}[\Psi; \delta \Psi]  & = \int_{\Sigma} (\Theta ^{\mu}[ \Phi ; I_{\hat{\chi}}\delta  \Phi] + L[\Phi] I_{\hat{\chi}}\Delta_{\text{X}}^{\mu} +\partial_{\nu} \Pi^{\mu \nu}[\Psi ; I_{\hat{\chi}}\delta \Psi ]) d^{3}x_{\mu} \\
       &  = - \int_{\Sigma} \{ 2 E_{(g) \nu}^{\mu} \xi^{\nu} + E_{(A)}^{\mu} ( \lambda + A_{\alpha} \xi^{\alpha} ) \} d^{3}x_{\mu}\\
       & \simeq 0 .
  \end{split}
\end{equation}
Here, the symbol $ \simeq$ emphasize that the equality holds on-shell. Taking an exterior derivative of the extended simplectic form yields the extended symplectic form,
\begin{equation}\label{50}
  \tilde{\Omega} [\Psi; \delta \Psi, \delta \Psi]= \delta \tilde{\boldsymbol{\Theta}}[\Psi; \delta \Psi].
\end{equation}
A fundamental result of the covariant phase space method is that generator $H_{\chi}[\Phi]$ of an infinitesimal gauge transformation acting on the fields is defined by the variational formula
\begin{equation}\label{51}
  \delta H_{\chi}[\Phi]= I_{\hat{\chi}}\tilde{\Omega} [\Psi; \delta \Psi, \delta \Psi],
\end{equation}
so that its Poisson bracket with the fields is given by
\begin{equation}\label{52}
  I_{\hat{\chi}} \delta F = \{ F, H_{\chi} \}_{\text{P.B}},
\end{equation}
where $F$ is an arbitrary functional on extended phase space.\\
To find explicit form of the Hamiltonians $H_{\chi}[\Phi]$ and the algebra among them, we assume that $\xi^{\mu}$ and $\lambda$ in $\chi$ are test functions so that combined transformation generated by $\chi$ is independent of solution. Due to this assumption we have $\Delta_{\text{Y}} \doteq \delta \lambda \doteq 0$. Here the symbol $\doteq$ emphasize that equality holds when components of $\chi$ are test functions. In this case, Eq.\eqref{43} reduces to
\begin{equation}\label{53}
  \tilde{\boldsymbol{\Theta}}[ \mathcal{E}(\text{Y},\mathfrak{g}_{\lambda})^{*}\Psi ; \delta  \mathcal{E}(\text{Y},\mathfrak{g}_{\lambda})^{*}\Psi] \doteq \tilde{\boldsymbol{\Theta}}[ \Psi ; \delta  \Psi].
\end{equation}
Hence, the $\mathcal{S}$ Lie derivative of extended symplectic potential along $\hat{\chi}$ vanishes,
\begin{equation}\label{54}
  \mathcal{L}_{\hat{\chi}}\tilde{\boldsymbol{\Theta}}[ \Psi ; \delta  \Psi]= \delta I_{\hat{\chi}} \tilde{\boldsymbol{\Theta}}+I_{\hat{\chi}} \delta \tilde{\boldsymbol{\Theta}} \doteq 0.
\end{equation}
This implies that the Hamiltonian is simply given by
\begin{equation}\label{55}
\begin{split}
   H_{\chi}[\Phi] &  = - I_{\hat{\chi}} \tilde{\boldsymbol{\Theta}}[ \Psi ; \delta  \Psi] \\
     & = \int_{\Sigma} \{ 2 E_{(g) \nu}^{\mu} \xi^{\nu} + E_{(A)}^{\mu} ( \lambda + A_{\alpha} \xi^{\alpha} ) \} d^{3}x_{\mu},
\end{split}
\end{equation}
where Eq.\eqref{49} was used. The Hamiltonian $H_{\chi}[\Phi]$ is constructed out of linear combination of equations of motion and it vanishes on-shell.\\
Now we return to the Eq.\eqref{50} and try to simplify it. By substituting Eq.\eqref{40} into Eq.\eqref{50}, we have
\begin{equation}\label{56}
  \begin{split}
     \tilde{\Omega} [\Psi; \delta \Psi, \delta \Psi] &= \delta \int_{\sigma} \text{X}^{*} \tilde{\Theta} ^{\mu}[ \Psi ; \delta  \Psi] d^{3}x_{\mu} \\
       & = \int_{\sigma} \text{X}^{*} ( \delta \tilde{\Theta} ^{\mu}[ \Psi ; \delta  \Psi]+\pounds_{\Delta_{\text{X}}} \tilde{\Theta} ^{\mu}[ \Psi ; \delta  \Psi]) d^{3}x_{\mu} \\
       & = \Omega  [\Phi; \delta \Phi, \delta \Phi]+\int_{\Sigma} E_{\Phi} \delta \Phi \Delta_{\text{X}}^{\mu} d^{3}x_{\mu} \\
       & + \int_{\partial \Sigma} \{ \delta \Pi^{\mu \nu} [\Psi; \delta \Psi] + \pounds_{\Delta_{\text{X}}} \Pi^{\mu \nu} [\Psi; \delta \Psi] - L[\Phi] \Delta_{\text{X}}^{\mu} \Delta_{\text{X}}^{\nu}-2 \Delta_{\text{X}}^{[ \mu} \Theta^{\nu ]}[\Phi; \delta \Phi] \} d^{2}x_{\mu \nu}.
  \end{split}
\end{equation}
In the above equation $\Omega  [\Phi; \delta \Phi, \delta \Phi]$ is non-extended symplectic form and it is given by
\begin{equation}\label{57}
  \begin{split}
     \Omega  [\Phi; \delta \Phi, \delta \Phi] = \int_{\Sigma} \sqrt{-g} \biggl\{ & -h^{\alpha \beta} \nabla_{\alpha} h^{\mu}_{\beta} + \frac{1}{2} h^{\alpha \beta} \nabla^{\mu} h_{\alpha \beta} + \frac{1}{2} h^{\mu \alpha} \nabla_{\alpha}h + \frac{1}{2} h \nabla_{\alpha} h^{\mu \alpha} -\frac{1}{2} h \nabla^{\mu} h  \\
       & -16 \pi \left( \delta F^{\mu \nu} \delta A_{\nu} + \frac{1}{2} h F^{\mu \nu} \delta A_{\nu} \right) \biggr\} d^{3}x_{\mu},
  \end{split}
\end{equation}
where $h_{\mu \nu}= \delta g_{\mu \nu}$ and $h= g^{\mu \nu} h_{\mu \nu}$. Because the second term in the last equality in Eq.\eqref{57} vanishes on-shell then it is not important (in fact, extended symplectic form is conserved on-shell and the mentioned term do not play role).\\
The equations \eqref{51} and \eqref{52} imply that algebra among Hamiltonians can be obtained as follows:
\begin{equation}\label{58}
   \begin{split}
       \{ H_{\chi_{1}} , H_{\chi_{2}} \}_{\text{P.B}} & \doteq I_{\hat{\chi}_{2}} I_{\hat{\chi}_{1}} \tilde{\Omega} \\
        & \doteq \int_{\Sigma} \{ 2 E_{(g) \nu}^{\mu} [\xi_{1} , \xi_{2}]^{\nu} + E_{(A)}^{\mu} ( \pounds_{\xi_{1}}\lambda_{2}-\pounds_{\xi_{2}}\lambda_{1} + A_{\alpha} [\xi_{1} , \xi_{2}]^{\alpha} ) \} d^{3}x_{\mu} \\
        & \doteq H_{[\chi_{1},\chi_{2}]},
   \end{split}
\end{equation}
where the algebra among generators of combined transformations are
\begin{equation}\label{59}
  [\chi (\xi_{1}, \lambda_{1} ),\chi (\xi_{2}, \lambda_{2})]= \chi (\xi_{12}, \lambda_{12}),
\end{equation}
with
\begin{equation}\label{60}
  \xi_{12} = [\xi_{1} , \xi_{2}] , \hspace{1 cm} \lambda_{12} =\pounds_{\xi_{1}}\lambda_{2}-\pounds_{\xi_{2}}\lambda_{1} .
\end{equation}
In this way, we have shown that algebra among Hamiltonians is isomorphic to the algebra among generators of combined symmetry transformations.
\section{Surface-preserving symmetries}\label{S.VI}
As in gauge theory, we can distinguish between two types of transformations. In section \ref{S.V}, we considered combined transformations which are constructed out of diffeomorphisms and $U(1)$ gauge transformations. They simply relabel the points of spacetime manifold and internal space of $U(1)$ gauge field, and hence are pure gauge. Therefore, these combined transformations are null directions in the symplectic form. However, there is a different class of combined transformations. We can consider them so that they transform the reference frame $(\text{X}^{\mu}, \Xi)$. These transformations have nontrivial generators and we will refer to them as surface symmetries \cite{1}.\\
To begin, let's consider diffeomorphism part of a combined transformation. As we saw earlier $\text{X}$ is a map from an open set $U \subset \mathbb{R}^{4}$ into $\mathcal{M}$, i.e. $\text{X}: U \rightarrow \mathcal{M}$. A transformation of the reference system can be considered as $\text{Z}: U \rightarrow U$. Its action can be defined as \cite{1}
\begin{equation}\label{61}
  g_{\mu \nu} \rightarrow g_{\mu \nu} , \hspace{1 cm} A_{\mu} \rightarrow A_{\mu} , \hspace{1 cm} \text{X} \rightarrow \text{X} \circ \text{Z},
\end{equation}
so that it changes the labelling of the points, but keeps the dynamical fields unchanged. By expanding $\text{Z}$ as $\text{Z}= \text{I}+ w + \mathcal{O}(w^{2})$, where $w$ is a vector field on $U$, we can find the infinitesimal version of this transformation. In order to take $U(1)$ gauge part into account we can consider combined transformation $\varpi = (w,\tau)$, where $\tau$ is a scalar function on $U$. Analogous to $\chi$, $\varpi$ defines a vector $\hat{\varpi}$ on $\mathcal{S}$. Because $\mathcal{E}(\text{X},\Xi)^{*} g_{\mu \nu} = \text{X}^{*}g_{\mu \nu}$ then the pullback field $\text{X}^{*}g_{\mu \nu}$ transforms as
\begin{equation}\label{62}
  \text{X}^{*}g_{\mu \nu} \rightarrow \text{Z}^{*} \text{X}^{*} g_{\mu \nu}.
\end{equation}
Therefore, the action of $\hat{\varpi}$ on $\mathcal{E}(\text{X},\Xi)^{*} g_{\mu \nu}$ is given by the $\mathcal{S}$ Lie derivative,
\begin{equation}\label{63}
  \begin{split}
     \mathcal{L}_{\hat{\varpi}} \mathcal{E}(\text{X},\Xi)^{*} g_{\mu \nu}&  = I_{\hat{\varpi}} \delta \text{X}^{*} g_{\mu \nu}\\
       & = I_{\hat{\varpi}} \text{X}^{*} (\delta g_{\mu \nu}+ \pounds_{\Delta_{\text{X}}} g_{\mu \nu})\\
       & = \text{X}^{*} (I_{\hat{\varpi}} \delta g_{\mu \nu}+ \pounds_{I_{\hat{\varpi}} \Delta_{\text{X}}} g_{\mu \nu}),
  \end{split}
\end{equation}
while its action on $g_{\mu \nu}$ is trivial,
\begin{equation}\label{64}
  \mathcal{L}_{\hat{\varpi}} g_{\mu \nu}=0 \hspace{0.5 cm} \Longleftrightarrow \hspace{0.5 cm} I_{\hat{\varpi}} \delta g_{\mu \nu}=0.
\end{equation}
Hence, we have
\begin{equation}\label{65}
   \mathcal{L}_{\hat{\varpi}} \mathcal{E}(\text{X},\Xi)^{*} g_{\mu \nu}= \text{X}^{*} \pounds_{W} g_{\mu \nu},
\end{equation}
where
\begin{equation}\label{66}
  W^{\mu}= I_{\hat{\varpi}} \Delta_{\text{X}}^{\mu},
\end{equation}
is the pushforward of the vector field $w$ to a vector field on $\mathcal{M}$. Similar to Eq.\eqref{65}, we can define the action of $\hat{\varpi}$ on $\mathcal{E}(\text{X},\Xi)^{*} A_{\mu}$ as
\begin{equation}\label{67}
   \mathcal{L}_{\hat{\varpi}} \mathcal{E}(\text{X},\Xi)^{*} A_{\mu}= \text{X}^{*} (\pounds_{W} A_{\mu}+ \partial _{\mu} \eta),
\end{equation}
where $\eta$ is a scalar field on $\mathcal{M}$. Because the action of $\hat{\varpi}$ on $A_{\mu}$ is trivial,
\begin{equation}\label{68}
  \mathcal{L}_{\hat{\varpi}} A_{\mu}=0 \hspace{0.5 cm} \Longleftrightarrow \hspace{0.5 cm} I_{\hat{\varpi}} \delta A_{\mu}=0,
\end{equation}
then
\begin{equation}\label{69}
  \begin{split}
     \mathcal{L}_{\hat{\varpi}} \mathcal{E}(\text{X},\Xi)^{*} A_{\mu} & = I_{\hat{\varpi}} \delta \text{X}^{*} \{ A_{\mu}+ \partial_{\mu} \Xi \} \\
       & =I_{\hat{\varpi}} \text{X}^{*} \{ \delta A_{\mu}+ \partial_{\mu}\delta \Xi+ \pounds_{\Delta_{\text{X}}} A_{\mu}+ \partial_{\mu} \pounds_{\Delta_{\text{X}}} \Xi \} \\
       & = \text{X}^{*} \{ \pounds_{W} A_{\mu} + \partial_{\mu} (I_{\hat{\varpi}} \delta \Xi +\pounds_{W} \Xi ) \}.
  \end{split}
\end{equation}
By comparing Eq.\eqref{67} and Eq.\eqref{69}, we find that
\begin{equation}\label{70}
   I_{\hat{\varpi}} \delta \Xi = -\pounds_{W} \Xi+ \eta.
 \end{equation}
 This equation relates $\eta$ and $\tau$ and we can regard $I_{\hat{\varpi}} \delta \Xi$ as pushforward of the scalar field $\tau$ to a scalar field on $\mathcal{M}$.\\
Suppose that $\varpi$ is independent of the solution, so that $\delta \varpi =0$, then
\begin{equation}\label{71}
  \begin{split}
     0 & = (\delta w , \delta \tau)\\
       & = (\delta \text{X}^{*} I_{\hat{\varpi}} \Delta_{\text{X}} ,  \delta \text{X}^{*} I_{\hat{\varpi}} \delta \Xi) \\
       & = \text{X}^{*} (\delta W + \pounds_{\Delta_{\text{X}}} W , - \delta \pounds_{W} \Xi +\delta \eta +\pounds_{\Delta_{\text{X}}} \{-\pounds_{W} \Xi + \eta \} ).
  \end{split}
\end{equation}
From the above equation we can deduce that nontrivial variations of $W$ and $\eta$ are
\begin{equation}\label{72}
  \delta W = [W, \Delta_{\text{X}}] ,
\end{equation}
\begin{equation}\label{73}
  \delta \eta = \delta \pounds_{W} \Xi + \pounds_{\Delta_{\text{X}}} \pounds_{W} \Xi - \pounds_{\Delta_{\text{X}}} \eta ,
\end{equation}
respectively.\\
The combined transformation generated by $\varpi$ is a symmetry of the phase space and then it generates a Hamiltonian flow $\mathcal{H}_{\varpi}$ via
\begin{equation}\label{74}
\begin{split}
  \delta \mathcal{H}_{\varpi}[\Phi] &= I_{\hat{\varpi}}\tilde{\Omega} [\Psi; \delta \Psi, \delta \Psi] \\
  &= - \int_{\partial \Sigma} \biggl\{ \delta Q^{\mu \nu}(W,\eta) + \pounds_{\Delta_{\text{X}}} Q^{\mu \nu}(W,\eta) + 2 W^{[ \mu} \left( \Theta^{\nu ]}[\Phi, \delta \Phi] + I_{\hat{\varkappa}} \Theta^{\nu ]}[\Phi, \delta \Phi] \right)\\
  & \hspace{1.7 cm} +  4 W^{[ \mu} E_{(g) \alpha}^{\nu ]} \Delta_{\text{X}}^{\alpha}+ 2 W^{[ \mu} E_{(A)}^{\nu ]} \left(\delta \Xi + A_{\alpha}\Delta_{\text{X}}^{\alpha} +\pounds_{\Delta_{\text{X}}} \Xi  \right) \biggr\} d^{2}x_{\mu \nu} \\
  & \hspace{0.5 cm}- \int_{\Sigma} E_{\Phi} \delta \Phi W^{\mu} d^{3}x_{\mu} ,
\end{split}
\end{equation}
where $\varkappa = (\Delta_{\text{X}}, \delta \Xi + \pounds_{\Delta_{\text{X}}} \Xi)$ and
\begin{equation}\label{75}
  Q^{\mu \nu}(W,\eta)= - 2 \sqrt{-g} \{ \nabla^{[ \mu} W^{\nu ]} + 8 \pi F^{\mu \nu} (\eta + A_{\alpha} W^{\alpha})  \}
\end{equation}
is the Noether charge in Einstein-Maxwell theory. The surface symmetry algebra is generated through the Poisson bracket of the Hamiltonians $\mathcal{H}_{\varpi}$ for all possible surface-preserving symmetry generators $\varpi$. One can show that the Poisson bracket is given by
\begin{equation}\label{76}
  \begin{split}
     \{ \mathcal{H}_{\varpi_{1}} , \mathcal{H}_{\varpi_{2}}\}_{\text{P.B}} & = I_{\hat{\varpi}_{2}} I_{\hat{\varpi}_{1}} \tilde{\Omega} [\Psi; \delta \Psi, \delta \Psi] \\
     & =- \int_{\partial \Sigma} \biggl( Q^{\mu \nu}\left([W_{1},W_{2}],\pounds_{W_{1}}\eta_{2}-\pounds_{W_{2}}\eta_{1} \right) -2 L[\Phi] W_{1}^{[ \mu} W_{2}^{\nu ]}- 2 W_{2}^{[ \mu} I_{\hat{\varpi}_{1}} I_{\hat{\varkappa}} \Theta^{\nu ]}[\Phi, \delta \Phi]  \\
       & \hspace{1.7 cm}+ 2 W_{1}^{[ \mu} I_{\hat{\varpi}_{2}} I_{\hat{\varkappa}} \Theta^{\nu ]}[\Phi, \delta \Phi] + 4 W_{1}^{[ \mu} E_{(g) \alpha}^{\nu ]} W_{2}^{\alpha}-  4 W_{2}^{[ \mu} E_{(g) \alpha}^{\nu ]} W_{1}^{\alpha} \\
       & \hspace{1.7 cm} + 2 W{1}^{[ \mu} E_{(A)}^{\nu ]} \left( \eta_{2} + A_{\alpha}W_{2}^{\alpha}  \right)-2 W{2}^{[ \mu} E_{(A)}^{\nu ]} \left( \eta_{1} + A_{\alpha}W_{1}^{\alpha}  \right)  \biggr)d^{2}x_{\mu \nu}.
  \end{split}
\end{equation}
Now we introduce a local coordinate system $x^{\mu}=(x^{i}, \sigma^{A})$ where $x^{i} (i=0,1)$ denote coordinates normal to the surface $\partial \Sigma$ and $\sigma^{A}(A=2,3)$ denote coordinates tangential to the surface $\partial \Sigma$. In this way, we can
make a $2+2$ decomposition of the metric in a neighbourhood of $\partial \Sigma$,
\begin{equation}\label{77}
  ds^{2}= \gamma_{i j} dx^{i} dx^{j} + q_{A B} (d \sigma^{A} -\mathcal{U}^{A}_{i} dx^{i})(d \sigma^{B} -\mathcal{U}^{B}_{j} dx^{j}),
\end{equation}
where $q_{A B}$ is induced metric on $\partial \Sigma$, $\gamma_{i j}$ is a generalized lapse which defines the normal geometry and $\mathcal{U}^{A}_{i}$ is a generalized shift. Let $n^{(i)}=n^{(i)}_{i} dx^{i}$ be normals to the level sets of constant $x^{i}$. Let $n^{(i)}_{i}$ be entries of a $2 \times 2$ matrix, say $n$. Hence the generalized lapse can be written as
\begin{equation}\label{78}
  \gamma_{i j} = n^{(i)}_{i} \eta_{(i)(j)} n^{(j)}_{j},
\end{equation}
where $\eta_{(i)(j)} = \text{diag}(-1,+1)$ is a flat 2D normal metric. By demanding that the equation $n^{(i)} \cdot n_{(j)} =\delta^{(i)}_{(j)}$ to be held and inner product of $n^{(i)}$ in $(d \sigma^{A} -\mathcal{U}^{A}_{i} dx^{i})$ vanishes, one can express the normal vector fields as
\begin{equation}\label{79}
  n_{(i)}= (n^{-1})_{(i)}^{i} (\partial_{i} + \mathcal{U}^{A}_{i} \partial_{A}).
\end{equation}
We can decompose $W$ into its normal part $W_{\perp}= W_{\perp}^{i} \partial_{i}$ and tangential part $W_{\parallel}= W_{\parallel}^{A} \partial_{A}$ so that $W = W_{\perp}+W_{\parallel}$. Here we want to focus on the surface-preserving transformations. For such transformations we have $W_{\perp} =0$ on $\partial \Sigma$. Despite the fact that for this class of transformations the normal part of $W$ vanishes on $\partial \Sigma$ but its partial derivatives does not, i.e we could have $\partial _{j} W_{\perp}^{i} \neq 0 $ on $\partial \Sigma$. By these assumptions, the equation \eqref{74} will be reduced to
\begin{equation}\label{80}
\begin{split}
  \delta \mathcal{H}_{\varpi}[\Phi] &= - \int_{\partial \Sigma} \bigl\{ \delta Q^{\mu \nu}(W,\eta) + \pounds_{\Delta_{\text{X}}} Q^{\mu \nu}(W,\eta)\bigr\} d^{2}x_{\mu \nu} \\
  &= -  \int_{\partial \sigma}  \text{X}^{*}\bigl\{ \delta Q^{\mu \nu}(W,\eta) + \pounds_{\Delta_{\text{X}}} Q^{\mu \nu}(W,\eta)\bigr\} d^{2}x_{\mu \nu}\\
  & = -  \delta \int_{\partial \sigma}  \text{X}^{*} Q^{\mu \nu}(W,\eta) d^{2}x_{\mu \nu}.
\end{split}
\end{equation}
Therefore, we can extract the Hamiltonian flow associated with surface-preserving transformations,
\begin{equation}\label{81}
  \mathcal{H}_{\varpi}[\Phi] \hat{=} - \int_{\partial \Sigma} Q^{\mu \nu}(W,\eta) d^{2}x_{\mu \nu}.
\end{equation}
Here the symbol $\hat{=}$ indicates that the equality holds for surface-preserving transformations. In a similar way, from Eq.\eqref{76}, one can deduce that the algebra among Hamiltonians conjugate to surface-preserving transformations is given by
\begin{equation}\label{82}
  \{ \mathcal{H}_{\varpi_{1}} , \mathcal{H}_{\varpi_{2}}\}_{\text{P.B}} \hat{=} \mathcal{H}_{\varpi_{12}},
\end{equation}
where
\begin{equation}\label{83}
  \mathcal{H}_{\varpi_{12}}= - \int_{\partial \Sigma} Q^{\mu \nu}(W_{12},\eta_{12}) d^{2}x_{\mu \nu},
\end{equation}
with
\begin{equation}\label{84}
  W_{12}= [W_{1},W_{2}],\hspace{1 cm}\eta_{12}= \pounds_{W_{1}}\eta_{2}-\pounds_{W_{2}}\eta_{1}.
\end{equation}
Bi-normal to $\partial \Sigma $ can be expressed as $\boldsymbol{\epsilon}_{\mu \nu}= \varepsilon^{(i)(j)}n_{(i) \mu}n_{(j)\nu}$ so that it is normalized to $-2$, where $\varepsilon^{(i)(j)}$ is the antisymmetric Levi-Civita symbol with $\varepsilon^{(0)(1)}=1$. In this way, the Hamiltonian \eqref{81} can be written as
\begin{equation}\label{85}
\begin{split}
  \mathcal{H}_{\varpi}[\Phi] &\hat{=} - \int_{\partial \Sigma} Q^{\mu \nu}(W,\eta) \boldsymbol{\epsilon}_{\mu \nu} d^{2}\sigma \\
  & \hat{=} 2 \int_{\partial \Sigma} d^{2}\sigma \sqrt{q} \{ \varepsilon^{(i)(j)} \nabla_{n_{(i)}} (W \cdot n_{(j)}) - [n_{(0)},n_{(1)}]^{\alpha} W_{\alpha}- \frac{\hat{\textbf{F}} }{\sqrt{q}}(\eta +A_{\alpha}W^{\alpha})\},
\end{split}
\end{equation}
where $q= \text{det}(q_{AB})$ and $\hat{\textbf{F}}=-16 \pi \sqrt{q} n_{(0)}^{\mu} n_{(1)}^{\nu} F_{\mu \nu}$ is proportional to the normal component of electric field. The first two terms in integrand are the contribution of the pure gravity part in Lagrangian and the explicit form of them have been calculated in \cite{1}. The last term in integrand is the new one and comes from the Maxwell term in Lagrangian. We can rewrite Eq.\eqref{85} as follows:
\begin{equation}\label{86}
  \mathcal{H}_{\varpi}[\Phi] \hat{=} -2 \int_{\partial \Sigma} d^{2}\sigma \{ \hat{\textbf{H}}_{i}^{\hspace{1.5 mm} j} \partial_{j}W_{\perp}^{i}+ \hat{\textbf{G}}_{A} W_{\parallel}^{A} +\hat{\textbf{F}} \eta \},
\end{equation}
where
\begin{equation}\label{87}
  \hat{\textbf{H}}_{i}^{\hspace{1.5 mm} j} = \frac{\sqrt{q}}{\text{det}(n)} \gamma_{ik} \varepsilon^{kj},
\end{equation}
\begin{equation}\label{88}
  \hat{\textbf{G}}_{A} = \frac{\sqrt{q}}{\text{det}(n)} q_{AB} \varepsilon^{ij} (\partial_{i}\mathcal{U}_{j}^{B} +\mathcal{U}_{i}^{C} \partial_{C}\mathcal{U}_{j}^{B})+ \hat{\textbf{F}} A_{A} .
\end{equation}
It is easy to check that $\hat{\textbf{H}}_{i}^{\hspace{1.5 mm} j}$ is traceless and $\text{det} (\hat{\textbf{H}})= -\text{det}(q)$. The algebra among $\hat{\textbf{H}}_{i}^{\hspace{1.5 mm} j}$, $\hat{\textbf{G}}_{A}$ and $\hat{\textbf{F}}$ can be extracted from Eq.\eqref{82}:
\begin{eqnarray}
  -2 \{\hat{\textbf{H}}_{i}^{\hspace{1.5 mm} j}(\sigma) , \hat{\textbf{H}}_{k}^{\hspace{1.5 mm} l}(\sigma^{\prime}) \}_{\text{P.B}} &=& (\delta^{l}_{i} \hat{\textbf{H}}_{k}^{\hspace{1.5 mm} j}-\delta^{j}_{k} \hat{\textbf{H}}_{i}^{\hspace{1.5 mm} l}) \delta^{2} (\sigma-\sigma^{\prime}) , \label{89} \\
  -2 \{\hat{\textbf{G}}_{A}(\sigma ) , \hat{\textbf{G}}_{B}(\sigma^{\prime}) \}_{\text{P.B}} &=& \hat{\textbf{G}}_{A}(\sigma^{\prime}) \partial_{B} \delta^{2} (\sigma-\sigma^{\prime})-\hat{\textbf{G}}_{B}(\sigma) \partial_{A}^{\prime} \delta^{2} (\sigma-\sigma^{\prime}), \label{90} \\
  -2 \{\hat{\textbf{H}}_{i}^{\hspace{1.5 mm} j}(\sigma), \hat{\textbf{G}}_{A}(\sigma^{\prime}) \}_{\text{P.B}} &=& \hat{\textbf{H}}_{i}^{\hspace{1.5 mm} j}(\sigma^{\prime}) \partial_{A} \delta^{2} (\sigma-\sigma^{\prime}), \label{91} \\
  -2 \{ \hat{\textbf{F}}(\sigma), \hat{\textbf{G}}_{A}(\sigma^{\prime})\}_{\text{P.B}} &=& \hat{\textbf{F}}(\sigma^{\prime}) \partial_{A} \delta^{2} (\sigma-\sigma^{\prime}),   \label{92} \\
  -2 \{\hat{\textbf{H}}_{i}^{\hspace{1.5 mm} j}(\sigma) ,  \hat{\textbf{F}}(\sigma^{\prime})\}_{\text{P.B}} &=& 0, \label{93} \\
  -2 \{\hat{\textbf{F}}(\sigma) ,  \hat{\textbf{F}}(\sigma^{\prime})\}_{\text{P.B}} &=& 0 . \label{94}
\end{eqnarray}
The components of $\hat{\textbf{G}}$ act as generators of tangential diffeomorphisms and $\hat{\textbf{H}}_{i}^{\hspace{1.5 mm} j}$ and $\hat{\textbf{F}}$ transform as scalars under diffeomorphism, while $\hat{\textbf{H}}_{i}^{\hspace{1.5 mm} j}$ itself generates a local $\mathfrak{sl}(2,\mathbb{R})$ algebra. Also, $\hat{\textbf{F}}$ itself generates a local $\mathfrak{u}(1)$ algebra. Therefore, the group of surface-preserving symmetries is semi-direct sum of 2-dimensional diffeomorphism group on $\partial \Sigma$ with $SL(2,\mathbb{R})$ and $U(1)$.\\
Because $\hat{\textbf{H}}_{i}^{\hspace{1.5 mm} j}$ is traceless, and hence $\hat{\textbf{H}}_{1}^{\hspace{1.5 mm} 1}=-\hat{\textbf{H}}_{0}^{\hspace{1.5 mm} 0}$, we can introduce local $\mathfrak{sl}(2,\mathbb{R})$ generators:
\begin{equation}\label{95}
  \hat{\textbf{K}}_{\pm}=2 (\hat{\textbf{H}}_{0}^{\hspace{1.5 mm} 1} \pm \hat{\textbf{H}}_{1}^{\hspace{1.5 mm} 0}), \hspace{1 cm} \hat{\textbf{K}}_{0}=-4 \hat{\textbf{H}}_{0}^{\hspace{1.5 mm} 0},
\end{equation}
so that they satisfy the $\mathfrak{sl}(2,\mathbb{R})$ commutation relations
\begin{equation}\label{96}
  \{\hat{\textbf{K}}_{+} ,  \hat{\textbf{K}}_{-}\}_{\text{P.B}}= 2 \hat{\textbf{K}}_{0}, \hspace{1 cm} \{\hat{\textbf{K}}_{\pm} ,  \hat{\textbf{K}}_{0}\}_{\text{P.B}}= 2 \hat{\textbf{K}}_{\mp}.
\end{equation}
Hence, similar to the pure gravity case \cite{1}, the Casimir of $SL(2,\mathbb{R})$ is the area element. Here, generators of tangential diffeomorphisms $\hat{\textbf{G}}_{A}$ differ from the one in the pure gravity (See Ref.\cite{1}) by an additional term $\hat{\textbf{F}} A_{A}$ which comes from the Maxwell term in the action.
\section{Conclusion}
Dynamical fields in the Einstein-Maxwell theory are spacetime metric $g_{\mu \nu}$ and the $U(1)$ gauge field $A_{\mu}$. This theory is not only covariant under diffeomorphisms but also is covariant under $U(1)$ gauge transformations. There are some evidence that these two transformations should be combined \cite{6,8,9,10,11}. For this purpose, we have introduced a combined map as $\mathcal{E}(\text{Y}, \mathfrak{g}_{\lambda})$ so that it induces a transformation $\mathcal{E}(\text{Y}, \mathfrak{g}_{\lambda}) ^{*} \Phi= \text{Y}^{*} \mathfrak{g}_{\lambda}^{*} \Phi$ on dynamical fields. We have considered covariant phase space method of obtaining conserved charges in classical field theories, where conserved charges can be extracted from symplectic form which is the exterior derivative of symplectic potential on phase space. We inferred that symplectic potential is not invariant under combined transformation, i.e.  under both diffeomorphism and $U(1)$ gauge transformation. In order to deal with this problem, we followed Donnelly and Freidel proposal and introduced new fields $\text{X}$ and $\Xi$ via replacing $\Phi \rightarrow \mathcal{E}(\text{X}, \mathfrak{g}_{\Xi})^{*} \Phi$. In this way, the phase space and consequently symplecic potential have been extended. Introduced fields under the action of combined transformation behave so that the extended symplectic potential is invariant under combined transformations at least on-shell or for transformations which do not depend on the solutions (see equations \eqref{41}, \eqref{42} and \eqref{43}). The introduction of combined transformation caused that a vector on phase space to be extended so that it contains a $U(1)$ part as well as the part induced by diffeomorphism (see section \ref{S.V}). Consequently, we showed that Hamiltonian $H_{\chi}[\Phi]$ conjugate to generators $\chi=(\xi, \lambda)$ of combined symmetry transformations is given by \eqref{55}. The Hamiltonian $H_{\chi}[\Phi]$ is constructed out of linear combination of equations of motion and it vanishes on-shell. In fact, the equation \eqref{55} implies that there exist five constraints in the Einstein-Maxwell theory (four of them come from diffeomorphism covariance and the other comes from covariance under $U(1)$ gauge transformation). The algebra among Hamiltonians has been calculated (see Eq.\eqref{58}). It was shown that the obtained algebra is closed and is isomorphic to the algebra among generators of combined symmetry transformations.\\
As in gauge theory, we can distinguish between two types of transformations. In section \ref{S.V}, we considered combined transformations which are constructed out of diffeomorphisms and $U(1)$ gauge transformations. They simply relabel the points of spacetime manifold and internal space of $U(1)$ gauge field, and hence are pure gauge. Therefore, these combined transformations are null directions in the symplectic form. However, there is a different class of combined transformations and we have investigated them in section \ref{S.VI}. In fact, they transform the reference frame $(\text{X}^{\mu}, \Xi)$ and hence their action on dynamical fields $\Phi$ are trivial (see Eq.\eqref{64} and Eq.\eqref{68}) while they act on introduced fields non-trivially (see Eq.\eqref{66} and Eq.\eqref{70}). Then we applied covariant space method to find Hamiltonians generating surface-preserving symmetries on a spacelike codimension two surface $\partial \Sigma$. These Hamiltonians have been obtained as in Eq.\eqref{86}. The algebra among Hamiltonians generating surface-preserving symmetries, Eq.\eqref{82}, implies the algebra given by the equations \eqref{89}-\eqref{94}. The algebra \eqref{89}-\eqref{94}, implies that the components of $\hat{\textbf{G}}$ act as generators of tangential diffeomorphisms and $\hat{\textbf{H}}_{i}^{\hspace{1.5 mm} j}$ and $\hat{\textbf{F}}$ transform as scalars under diffeomorphism, while $\hat{\textbf{H}}_{i}^{\hspace{1.5 mm} j}$ itself generates a local $\mathfrak{sl}(2,\mathbb{R})$ algebra. Also, $\hat{\textbf{F}}$ itself generates a local $\mathfrak{u}(1)$ algebra. Therefore, the group of surface-preserving symmetries is semi-direct sum of 2-dimensional diffeomorphism group on $\partial \Sigma$ with $SL(2,\mathbb{R})$ and $U(1)$. Here, similar to the pure gravity case \cite{1}, the Casimir of $SL(2,\mathbb{R})$ is the area element.
\section{Acknowledgments}
The work of Hamed Adami has been financially supported by Research Institute for Astronomy Astrophysics of Maragha (RIAAM).

\appendix
\section{Pullback and its Variation}\label{APP.A}
Following Ref.\cite{1}, we want to introduce a variational formula for pullback of a generic tensor density $\mathcal{T}$. To find this formula we first establish it on a scalar, a contravariant vector, a covariant vector and then on a scalar density of weight $+1$, respectively. One can then extend the formula to a generic tensor density by the product rule.\\
Let $\text{Y} : \mathcal{M} \rightarrow \mathcal{M}$ be a diffeomorphism of spacetime and $\text{Y}^{*} \mathcal{T}$ denote the pullback under diffeomorphism. The action of pullback $\text{Y}^{*}$ on a scalar field $f: \mathcal{M} \rightarrow \mathbb{R}$ is given by
\begin{equation}\label{A1}
  \text{Y}^{*}f(x) = f(\text{Y}).
\end{equation}
By taking variation from Eq.\eqref{A1}, we have
\begin{equation}\label{A2}
\begin{split}
   \delta \text{Y}^{*}f(x) & = (\delta f)(\text{Y})+\delta \text{Y}^{\alpha^{\prime}}\partial_{\alpha^{\prime}} f(\text{Y}) \\
     & = \text{Y}^{*}( \delta f + \Delta_{\text{Y}}^{\alpha} \partial_{\alpha} f) \\
     & = \text{Y}^{*}( \delta f + \pounds _{\Delta_{\text{Y}}} f)
\end{split}
\end{equation}
where $ \Delta_{\text{Y}} = \delta \text{Y} \circ \text{Y}^{-1}$ was used. Here we introduce $\{\alpha^{\prime},... \}$ indices on $\text{Y}^{\alpha^{\prime}}$ to distinguish them from $\{ \alpha ,... \}$ indices on $x^{\alpha}$. The pullback map can be applied to
contravariant vectors by defining $\text{Y}^{*} = (\text{Y}^{-1})_{*}$ on contravariant vectors then for a contravariant vector $v^{\alpha}(x)$ we have
\begin{equation}\label{A3}
  \text{Y}^{*}v^{\alpha}(x) = \frac{\partial x^{\alpha}}{\partial \text{Y}^{\alpha^{\prime}}} v^{\alpha^{\prime}}(\text{Y}).
\end{equation}
Therefore, the variation of $\text{Y}^{*}v^{\alpha}(x)$ can be written as
\begin{equation}\label{A4}
  \begin{split}
     \delta \text{Y}^{*}v^{\alpha}(x) & =\delta \left( \frac{\partial x^{\alpha}}{\partial \text{Y}^{\alpha^{\prime}}} \right) v^{\alpha^{\prime}}(\text{Y})+ \frac{\partial x^{\alpha}}{\partial \text{Y}^{\alpha^{\prime}}} \left[ (\delta v^{\alpha^{\prime}})(\text{Y}) + \delta \text{Y}^{\beta^{\prime}} \partial_{\beta^{\prime}} v^{\alpha^{\prime}}(\text{Y}) \right] \\
       & =  \frac{\partial x^{\alpha}}{\partial \text{Y}^{\alpha^{\prime}}} \left[ (\delta v^{\alpha^{\prime}})(\text{Y}) + \delta \text{Y}^{\beta^{\prime}} \partial_{\beta^{\prime}} v^{\alpha^{\prime}}(\text{Y})-  v^{\beta^{\prime}}(\text{Y}) \partial_{\beta^{\prime}} \delta \text{Y}^{\alpha^{\prime}} \right] \\
       & = \text{Y}^{*} (\delta v^{\alpha}+ \pounds_{\Delta_{\text{Y}}} v^{\alpha}).
  \end{split}
\end{equation}
Also, for a covariant vector $w_{\alpha}$ we have:
\begin{equation}\label{A5}
  \begin{split}
     \delta \text{Y}^{*}w_{\alpha}(x) & =\delta \left( \frac{ \partial \text{Y}^{\alpha^{\prime}}}{\partial x^{\alpha}} w_{\alpha^{\prime}}(\text{Y}) \right) \\
       & =\delta \left( \frac{ \partial \text{Y}^{\alpha^{\prime}}}{\partial x^{\alpha}} \right) w_{\alpha^{\prime}}(\text{Y})+ \frac{ \partial \text{Y}^{\alpha^{\prime}}}{\partial x^{\alpha}} \left[ (\delta w_{\alpha^{\prime}})(\text{Y}) + \delta \text{Y}^{\beta^{\prime}} \partial_{\beta^{\prime}} w_{\alpha^{\prime}}(\text{Y}) \right] \\
       & =  \frac{ \partial \text{Y}^{\alpha^{\prime}}}{\partial x^{\alpha}} w_{\alpha^{\prime}}(\text{Y}) \left[ (\delta w_{\alpha^{\prime}})(\text{Y}) + \delta \text{Y}^{\beta^{\prime}} \partial_{\beta^{\prime}} w_{\alpha^{\prime}}(\text{Y})+  w_{\beta^{\prime}}(\text{Y}) \partial_{\alpha^{\prime}} \delta \text{Y}^{\beta^{\prime}} \right] \\
       & = \text{Y}^{*} (\delta w_{\alpha}+ \pounds_{\Delta_{\text{Y}}} w_{\alpha}).
  \end{split}
\end{equation}
Square root of metric determinant $\sqrt{-g}$ is a scalar density of weight $+1$ and the action of pullback on $\sqrt{-g(x)}$ is given by
\begin{equation}\label{A6}
  \text{Y}^{*} \sqrt{-g(x)}= \left|\frac{ \partial \text{Y}}{\partial x} \right| \sqrt{-g(\text{Y})},
\end{equation}
where $\left|\frac{ \partial \text{Y}}{\partial x} \right|= \text{det} \left( \frac{ \partial \text{Y}^{\alpha^{\prime}}}{\partial x^{\alpha}}\right)$, then
\begin{equation}\label{A7}
  \begin{split}
     \delta \text{Y}^{*} \sqrt{-g}& = \delta \left|\frac{ \partial \text{Y}}{\partial x} \right| \sqrt{-g(\text{Y})}+ \left|\frac{ \partial \text{Y}}{\partial x} \right| \left( \delta\sqrt{-g} (\text{Y}) + \delta \text{Y}^{\alpha^{\prime}} \partial_{\alpha^{\prime}} \sqrt{-g(\text{Y})} \right)\\
       & = \left|\frac{ \partial \text{Y}}{\partial x} \right| \left( \delta\sqrt{-g} (\text{Y}) + \delta \text{Y}^{\alpha^{\prime}} \partial_{\alpha^{\prime}} \sqrt{-g(\text{Y})}+ \sqrt{-g(\text{Y})} \partial_{\alpha^{\prime}} \delta \text{Y}^{\alpha^{\prime}} \right) \\
       & =\text{Y}^{*} ( \delta \sqrt{-g}+ \pounds _{\Delta_{\text{Y}}} \sqrt{-g} ).
  \end{split}
\end{equation}
Considering equations \eqref{A2}, \eqref{A4}, \eqref{A5} and \eqref{A7}, by virtue of the product rule, we can state that for a generic tensor density $\mathcal{T}$ we have
\begin{equation}\label{A8}
  \delta \text{Y}^{*}\mathcal{T}= \text{Y}^{*} ( \delta \mathcal{T}+ \pounds _{\Delta_{\text{Y}}} \mathcal{T} ).
\end{equation}
\section{Derivation of some useful equations}\label{APP.B}
\textbf{First equation.} Given $\chi_{1}=(\xi_{1}, \lambda_{1}) $ and $\chi_{2}=(\xi_{2}, \lambda_{2}) $. From Eq.\eqref{46}, we have $I_{\hat{\chi}_{1}} \Delta_{\text{X}}^{\mu} = -\xi_{1}^{\mu}$ and $I_{\hat{\chi}_{2}} \Delta_{\text{X}}^{\mu} = -\xi_{2}^{\mu}$. Consider
\begin{equation*}
  \begin{split}
      I_{\hat{\chi}_{1}} I_{\hat{\chi}_{2}}[\Delta_{\text{X}}, \Delta_{\text{X}}]^{\mu} &= I_{\hat{\chi}_{1}} ([I_{\hat{\chi}_{2}} \Delta_{\text{X}}, \Delta_{\text{X}}]^{\mu}-[\Delta_{\text{X}}, I_{\hat{\chi}_{2}}\Delta_{\text{X}}]^{\mu})  \\
       & = 2 I_{\hat{\chi}_{1}} [\Delta_{\text{X}}, \xi_{2}]^{\mu} \\
       & = -2  [\xi_{1}, \xi_{2}]^{\mu} \\
       & = -2 (\xi_{1}^{\nu} \nabla_{\nu} \xi_{2}^{\mu}-\xi_{2}^{\nu} \nabla_{\nu} \xi_{1}^{\mu})\\
       & = -2 (I_{\hat{\chi}_{1}} \Delta_{\text{X}}^{\nu} \nabla_{\nu} I_{\hat{\chi}_{2}} \Delta_{\text{X}}^{\mu}-I_{\hat{\chi}_{2}} \Delta_{\text{X}}^{\nu} \nabla_{\nu} I_{\hat{\chi}_{1}} \Delta_{\text{X}}^{\mu})\\
       & = 2 I_{\hat{\chi}_{1}} I_{\hat{\chi}_{2}} \Delta_{\text{X}}^{\nu} \nabla_{\nu} \Delta_{\text{X}}^{\mu},
  \end{split}
\end{equation*}
then we arrive at
\begin{equation}\label{B1}
  \frac{1}{2} [\Delta_{\text{X}}, \Delta_{\text{X}}]^{\mu}= \Delta_{\text{X}}^{\nu} \nabla_{\nu} \Delta_{\text{X}}^{\mu}.
\end{equation}
\textbf{Second equation.} Using the following equation
\begin{equation*}
   -\frac{1}{2} I_{\hat{\chi}_{1}} I_{\hat{\chi}_{2}}[\Delta_{\text{X}}, \Delta_{\text{X}}]^{\mu}= [\xi_{1}, \xi_{2}]^{\mu},
\end{equation*}
we can write
\begin{equation*}
\begin{split}
   -\frac{1}{2} I_{\hat{\chi}_{1}} I_{\hat{\chi}_{2}}\pounds_{[\Delta_{\text{X}}, \Delta_{\text{X}}]} & = \pounds_{\xi_{1}}\pounds_{\xi_{2}}-\pounds_{\xi_{2}}\pounds_{\xi_{1}} \\
     & = \pounds_{I_{\hat{\chi}_{1}} \Delta_{\text{X}}}\pounds_{I_{\hat{\chi}_{2}} \Delta_{\text{X}}}-\pounds_{I_{\hat{\chi}_{2}} \Delta_{\text{X}}}\pounds_{I_{\hat{\chi}_{1}} \Delta_{\text{X}}} \\
     &= -I_{\hat{\chi}_{1}} I_{\hat{\chi}_{2}} \pounds_{\Delta_{\text{X}}}\pounds_{\Delta_{\text{X}}},
\end{split}
\end{equation*}
then
\begin{equation}\label{B2}
  \pounds_{\Delta_{\text{X}}}\pounds_{\Delta_{\text{X}}}= \pounds_{\frac{1}{2} [\Delta_{\text{X}}, \Delta_{\text{X}}]}.
\end{equation}
\textbf{Third equation.}
\begin{equation*}
  \begin{split}
     I_{\hat{\chi}_{1}} I_{\hat{\chi}_{2}} \pounds_{\Delta_{\text{X}}}\Delta_{\text{X}}^{\mu} & = \pounds_{I_{\hat{\chi}_{2}} \Delta_{\text{X}}} I_{\hat{\chi}_{1}}\Delta_{\text{X}}^{\mu}-\pounds_{I_{\hat{\chi}_{1}}\Delta_{\text{X}}}I_{\hat{\chi}_{2}}\Delta_{\text{X}}^{\mu} \\
       & = 2 [\xi_{2},\xi_{1}]^{\mu} \\
       & = I_{\hat{\chi}_{1}} I_{\hat{\chi}_{2}}[\Delta_{\text{X}}, \Delta_{\text{X}}]^{\mu},
  \end{split}
\end{equation*}
then
\begin{equation}\label{B3}
  \pounds_{\Delta_{\text{X}}}\Delta_{\text{X}}=[\Delta_{\text{X}}, \Delta_{\text{X}}].
\end{equation}
\textbf{Forth equation.} Using nilpotency $\delta^{2}=0$, we have
\begin{equation*}
  \begin{split}
     0 & = \delta \delta \mathcal{E}(\text{X}, \mathfrak{g}_{\Xi})^{*} \Phi \\
       & = \delta \delta \text{X}^{*} (\Phi + \delta_{\Phi}^{A} \partial_{\mu} \Xi)\\
       & = \delta \text{X}^{*} (\delta \Phi + \delta_{\Phi}^{A} \partial_{\mu} \delta \Xi + \pounds_{\Delta_{\text{X}}} \Phi + \delta_{\Phi}^{A} \partial_{\mu} \pounds_{\Delta_{\text{X}}}\Xi )\\
       & = \text{X}^{*} ( \delta\pounds_{ \Delta_{\text{X}}} \Phi + \delta_{\Phi}^{A} \partial_{\mu} \delta \pounds_{\Delta_{\text{Y}}}\Xi + \pounds_{\Delta_{\text{X}}} \delta \Phi + \delta_{\Phi}^{A} \partial_{\mu} \pounds_{\Delta_{\text{X}}} \delta \Xi + \pounds_{\Delta_{\text{X}}} \pounds_{\Delta_{\text{X}}} \Phi + \delta_{\Phi}^{A} \partial_{\mu} \pounds_{\Delta_{\text{X}}} \pounds_{\Delta_{\text{X}}}\Xi ) \\
       & = \text{X}^{*} \{\pounds_{\delta \Delta_{\text{X}}}(\Phi+\delta_{\Phi}^{A} \partial_{\mu} \Xi)+\pounds_{\frac{1}{2} [\Delta_{\text{X}}, \Delta_{\text{X}}]}(\Phi+\delta_{\Phi}^{A} \partial_{\mu} \Xi) \},
  \end{split}
\end{equation*}
then
\begin{equation}\label{B4}
  \delta \Delta_{\text{X}}= -\frac{1}{2} [\Delta_{\text{X}}, \Delta_{\text{X}}].
\end{equation}

\end{document}